\documentclass{article}
\usepackage{epsf}
\usepackage{emulateapj}
\usepackage{apjfonts}
\usepackage{flushrt}

\lefthead{Churazov et al.}
\righthead{Merging in the Centaurus Cluster}
\slugcomment{\em Submitted to The Astrophysical Journal}

\def\kms{~km~s$^{-1}${}}
\def\gax    {{_>\atop^{\sim}}}
\def\deg{$^{\circ}$}

\tighten

\begin{document}

\title{Evidence for Merging in the Centaurus Cluster}

\author{E.Churazov, M.Gilfanov}
\affil{MPI fuer Astrophysik,
Karl-Schwarzschild-Str.1, 85740 Garching bei Munchen, Germany;  \\
Space Research Institute (IKI), Profsouznaya 84/32, Moscow 117810, 
Russia}

\and

\author{W.Forman, C.Jones}
\affil{Harvard-Smithsonian Center for Astrophysics, 60 Garden St., Cambridge, MA 02138}

\begin{abstract}
	We present a two dimensional map of the gas temperature
distribution in the Centaurus cluster, based on ASCA observations
derived using a novel approach to account for the energy dependent
point spread function. Along with a cool region, centered on NGC4696,
asymmetric temperature variations of moderate amplitude are
detected. The hottest region roughly coincides with the position of
the second brightest galaxy NGC4709, known to be the dominant galaxy
of one of the subgroups in the Centaurus cluster.  ROSAT images show
faint surface brightness emission also centered on this galaxy. The
imaging and spectral results suggest that a subcluster, centered on
NGC4709 is merging with the main cluster centered on NGC4696, in
agreement with the earlier suggestion by \cite{lcd86} that these two
systems are located at the same distance despite their different
line-of-sight velocities.
\end{abstract}


\section{INTRODUCTION}
	
The Centaurus cluster is a nearby (mean velocity of the dominant
component is about 3000 \kms, e.g. Lucey, Currie, \& Dickens 1986) X-ray bright cluster
with a 2--10 keV luminosity of $7\times 10^{43}$~ergs~s$^{-1}$ (Jones
\& Forman 1978). The cluster X--ray emission is strongly peaked on the
cD galaxy NGC4696, indicating a cooling flow with a moderate mass
deposition rate of $\sim$ 30--50 $M_\odot$ (e.g. Matilsky, Jones \&
Forman 1985; Allen \& Fabian 1994; White, Jones \& Forman 1997).

The cluster was intensively studied in X-rays with imaging instruments
by Einstein (Matilsky, Jones \& Forman 1985), ROSAT (Allen \& Fabian
1994), and ASCA (Fukazawa et al.\ 1994).  Rather smooth X-ray
isophotes, having a slightly elliptical shape (Allen \& Fabian 1994),
and the presence of a cooling flow suggest that the Centaurus cluster
is a relatively relaxed system. However, measurements of line-of-sight
velocities indicate the presence of two distinct components (Lucey et al.
1986), one centered on NGC4696 and another
dominated by NGC4709, but having a factor of 2.5 fewer galaxies. These
two components, denoted as Cen 30 and Cen 45, have mean velocities and
dispersions of ${\rm v}=3041$\kms, $\sigma=586$\kms~ and ${\rm v}=4570$\kms,
$\sigma= 280$\kms~ respectively (Lucey et al.
1986). Although this may be the simple projection of two spatially
separated groups, \cite{lcd86} argued, on the basis of comparison of
luminosity functions, color--magnitude relations and galaxian-radius
distributions, that both components are in fact located at the same
distance. This would imply that the main cluster, Cen 30, is in the
process of accreting the smaller group, Cen 45, at the present
time. We show below that the ASCA observations provide strong
support that the Centaurus cluster is undergoing a merger.

The structure of the paper is as follows: Section 2 briefly describes
the X--ray observations and data reduction procedure. In Section 3 we
discuss results of the analysis and present arguments in favor of an
ongoing merger of the Cen 30 and Cen 45 subclusters.
Section 4 summarizes the results.

\section{DATA ANALYSIS}

\begin{figure*}
\plotone{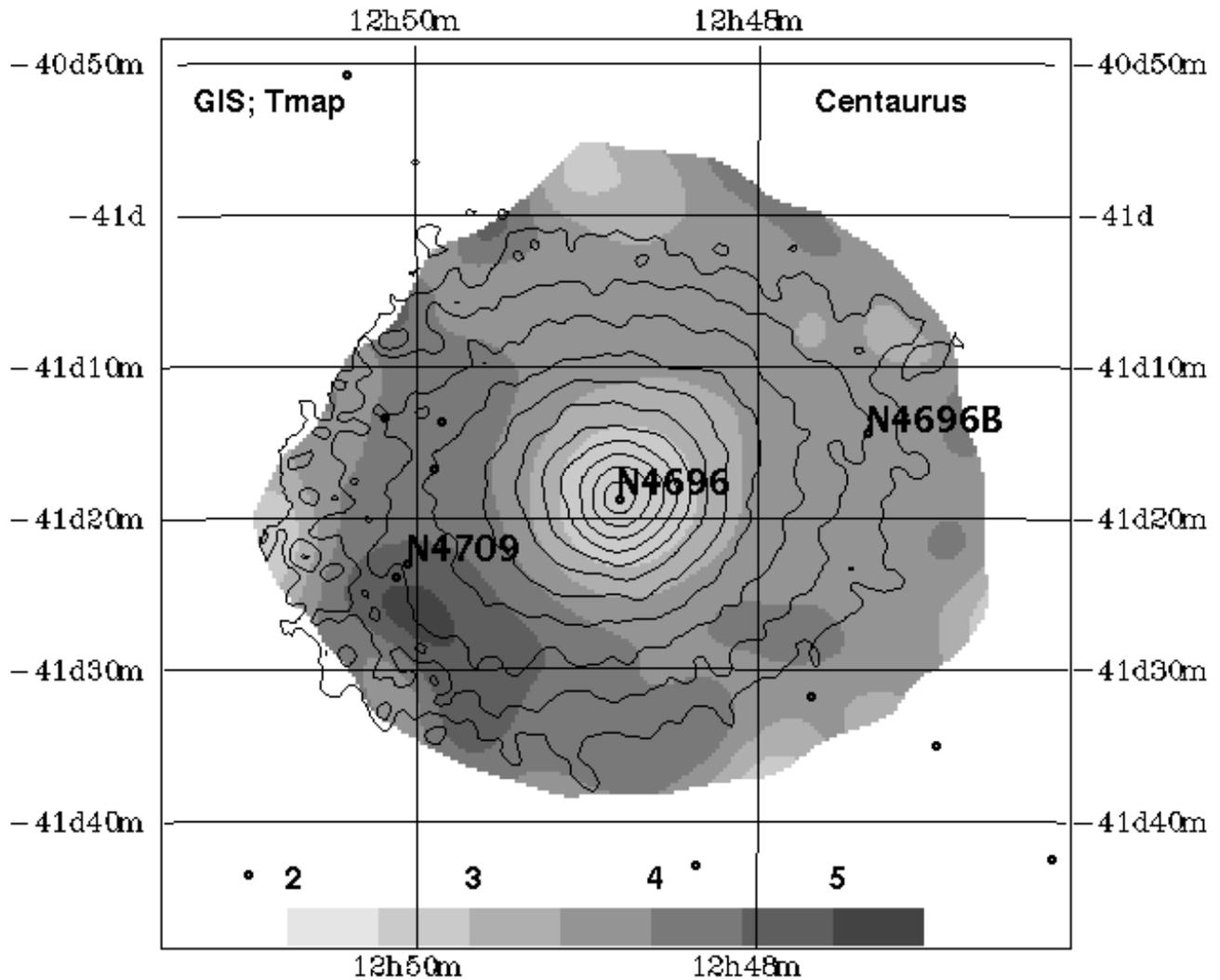}
\caption{
Greyscale image of the gas temperature distribution in the Centaurus
cluster (see also color Plate 1). The image was smoothed with a $2'$
Gaussian. Contours correspond to the 0.5--9 keV GIS surface brightness
image
corrected for background and vignetting. The surface brightness image
was
smoothed with a $0.5'$ Gaussian. Circles show the
positions of the brightest galaxies (brighter than m=14).
\label{gistmap}}
\end{figure*}

For the analysis presented below, we used three observations of the
Centaurus cluster taken with ASCA on 1993 June 30, July 5 
and 1995 July 19. 
These three observations have two slightly different pointing
directions and together cover the region extending from the SE to the
NW. Primarily GIS data were used for the analysis,
which has a larger field of view and more stable  characteristics
over 1993--1995, compared to the ASCA SIS. Standard screening has been applied
to the ASCA data with 
recommended values, forming a cleaned data set which was used for all
the analysis. Similar screening criteria have been applied to publicly
available observations of ``blank fields'', which were used for
background subtraction. The effective area was calculated using the
most recent version of the efficiency curves
(xrt\_ea\_v2\_0.fits). A contour map of the resulting background subtracted, vignetting
corrected, GIS image in the 0.5--2 keV energy band is shown
in Fig.~\ref{gistmap} (Plate 1). The circle coinciding with the
peak of the surface brightness corresponds to the dominant galaxy NGC4696.

To reconstruct the projected (two dimensional) gas temperature
distribution, we used the approach described in \cite{cgfj96}. This
method provides an efficient characterization of the shape of the
spectrum at a given position as a linear combination of template
spectra, corresponding to optically thin plasma emission models at two
different temperatures. For our analysis, we used the MEKAL model from
XSPEC (convolved with the efficiency of the mirror systems and the
appropriate ASCA detector) at temperatures of 2 and 8 keV, as template
spectra. For both template spectra, the heavy element abundance was
set to 0.5 of solar, the absorption column density to
$N_H=8\times10^{20}~{\rm cm}^{-2}$, and the redshift to 0.0104.  The
value of the temperature at a given position is calculated from a
simple combination of the best fit weights of the two template models
to the spectrum observed at this position. Due to the linear nature of
the method no actual fitting is required and, hence, the method is
computationally fast.  Although the limitations of this method are
significant (e.g., simple spectral forms must be assumed a priori),
this approach does provide a convenient method to examine two
dimensional spatial variations in gas temperature. Note that for a
single temperature spectrum, the accuracy of the method is sufficient
(for temperatures above $\sim$ 2 keV) as shown by the simulations in
\cite{cgfj96}.

The extended, energy dependent wings of the ASCA mirrors are known to
cause significant systematic errors in spectral analyses of extended
sources, usually leading to spurious temperature increases at large
radii from the cluster center where the scattered emission from the
cluster center contributes significantly to the total emission. We
developed an approach to correct for the effects of the PSF (for a
detailed description including a more rigorous derivation see Gilfanov
et al. 1998) which relies on the existence of a compact core in the
ASCA PSF. For the PSF model, we use publicly available GIS images of
Cygnus X--1 observed with different positional offsets.
The method is based
on the assumption that the ASCA PSF ($P$) can be represented as the
sum of two components:
\begin{equation} \label{psfma}
P=P_{core}+P_{wings}
\end {equation}
where $P_{core}$ corresponds to the rather compact ``core'' of the PSF and
$P_{wings}$  corresponds to the extended (large spatial scale) wings of
the PSF, in such a way that 
$\sum{P_{wings}}\ll 1$, i.e., the wings of the PSF contain
only a small fraction of the photons (we assume here that by definition
$\sum{P}\equiv 1$). The convolution of the true sky image ($S$) with the PSF
can be rewritten as 
\begin{equation} \label{convma}
D=S\otimes P_{core}+S\otimes P_{wings}
\end {equation}
where $D$ is the observed detector image. 
Provided that $P_{core}$ is sufficiently compact, the first term in the above
sum,  $S\otimes P_{core}$, would provide a resonable approximation to the sky
image $S$, since it is the
convolution of the true sky image with a relatively compact function. From
(\ref{convma}) it follows that 
\begin{equation} \label{scema}
S\otimes P_{core}=D-S\otimes P_{wings}
\end {equation}
Utilizing the fact that $\sum{P_{wings}}\ll 1$, one can obtain an
approximate solution, $Y$
for $S\otimes P_{core}$, by substituting the measured detector image
$D$ for $S$ in the right side of
equation (\ref{scema})
\begin{eqnarray}\label{scama}
Y = D-D\otimes P_{wings}=S\otimes P_{core}-(D-S)\otimes P_{wings} \nonumber \\
  = S\otimes P_{core} -(S\otimes P_{core}-S)\otimes P_{wings}  -
  S\otimes P_{wings}\otimes P_{wings} 
\end{eqnarray}
One can see that the two last terms in the above equation can be dropped,
assuming that the cluster image $S$ is ``smooth'' and $\sum{P_{wings}}\ll 1$.
Thus, 
\begin{eqnarray}\label{scbma}
Y\approx S\otimes P_{core} 
\end{eqnarray}

To practically implement  this method, we use the following
simple choice for  $P_{core}$ and $P_{wings}$:
\begin{eqnarray}\label{choicema}
P_{core}=P;~ P_{wings}=0, ~~{\rm if}~~ r < R_{core}=6' \nonumber \\
P_{core}=0;~ P_{wings}=P, ~~{\rm if}~~ r > R_{core}=6' \nonumber 
\end{eqnarray}
where $r$ is the distance from the center of the PSF.
The accuracy of the approximation in eq.~(\ref{scbma}) is determined by the
fraction of photons contained in the PSF wings,
$\frac{\sum{P_{wings}}}{\sum{P_{core}}}$, which in turn depends on the 
value of $R_{core}$ -- the larger is $R_{core}$, the better  the approximation. 
On the other hand, for larger values of  $R_{core}$,
the PSF core defined  above  becomes less compact and its energy and
position dependence becomes more important.
The choice of $R_{core}=6'$ was found to be adequate for clusters with
core radii larger than a few arcminutes. 

Equation~(\ref{scbma}) involves a two-dimensional convolution
($D\otimes P_{wings}$) of the observed detector image (in a given
energy band) with $P_{wings}$, which in turn depends on position and
energy. Straightforward convolution is a very time consuming
procedure, but a dramatic acceleration can be achieved via a
Monte-Carlo approach. For each detected photon, we choose $N=10-20$
random positions $x_i,y_i$ (distributed according to $P_{wings}$) and
subtract their contribution ($\sum{P_{wings}}/N$) from the
corresponding positions. The function $P_{wings}$ itself is represented as
a one-dimensional cumulative array, i.e. the two-dimensional array
$P_{wings}$ (of size $M\times M$) is represented as a one-dimensional
array of size $M^2$, whose $k$-th element is equal to the sum of all
elements from $P_{wings}$ with indices less than $k$.  Such a
representation of $P_{wings}$ allows one to generate random positions
(distributed like $P_{wings}$) with just a few arithmetical
operations, using the bisection method.

The simplicity of this method makes its implementation computationally
fast and it can be performed in each of the 1024 ASCA GIS energy
channels (or 512 SIS channels) on a fine ($\approx 15''$) angular
grid.  On the other hand, to decrease the systematic uncertainty
intrinsic to this method, the value of $R_{core}$ must be sufficiently
large. This restricts somewhat the applicability of this approach so
that it can be applied most accurately to large, generally nearby
clusters. In particular, the method has proven to work fairly well for
clusters of galaxies with core radii exceeding several arcmin (see
Gilfanov et al. 1998).

The resulting temperature map for the Centaurus cluster is shown in
Fig.~\ref{gistmap} (Plate 1). The statistical noise increases towards
the edges of the map and the spectral measurements are shown only for
those regions where the ratio of the upper limit on the temperature to
the lower limit is less than a factor of 1.5.  One can clearly see a
cool region centered on NGC4696 as well as asymmetric variations of
temperature. However, we wish to emphasize that these variations are
of moderate amplitude with the temperature ranging from $\sim$3--3.5
keV to $\sim$ 4--4.5 keV (excluding the cooling flow region). Note
that the hottest region to the SE is roughly coincident with the position of
NGC4709 which is the dominant galaxy of the Cen 45 subcluster
(Lucey et al. 1986).

\begin{figure*}
\plottwo{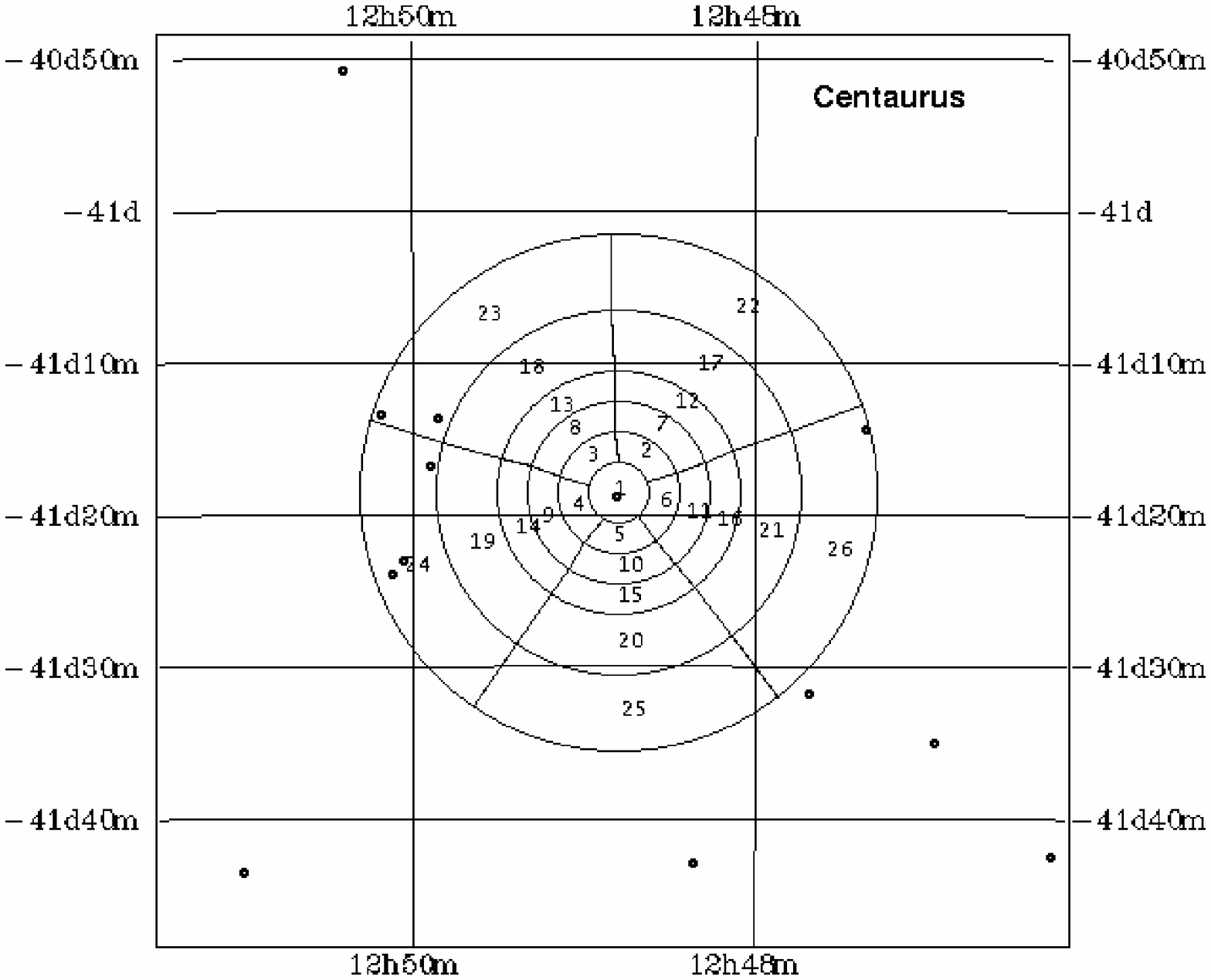}{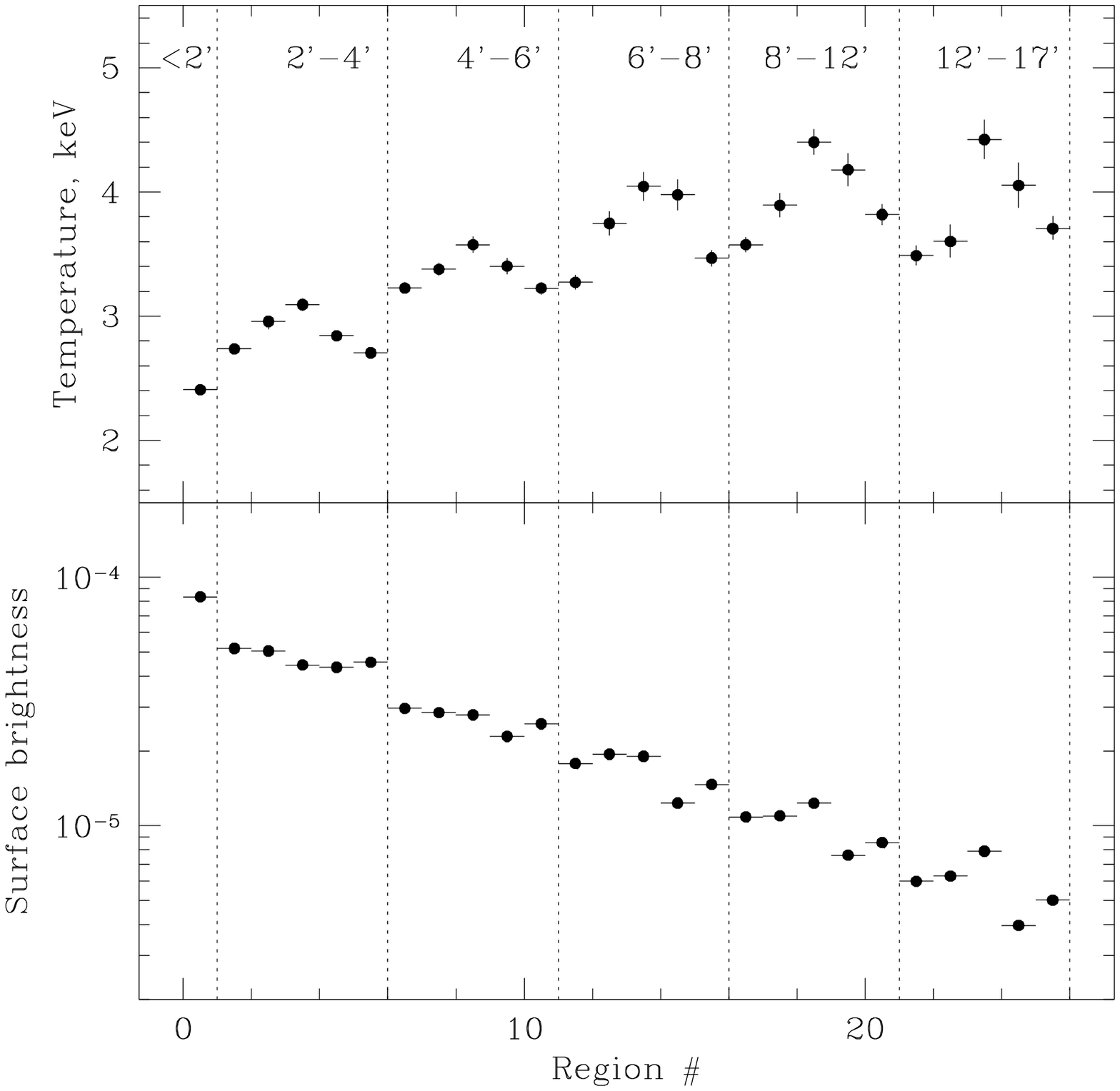}
\caption{{\bf Left:} The regions selected for spectral fitting. Solid
symbols
indicate the bright ($m<14$) galaxies in the field. {\bf Right:} The
temperature and surface brightness versus region
number. Errors indicate regions of 68 per cent confidence for a
single parameter of interest.
\label{gisreg}
\label{gsfit}
}
\end{figure*}

The same PSF correction method can be applied to spectra accumulated
from arbitrary regions. We have chosen a set of regions as shown in
Fig.~\ref{gisreg} and have accumulated GIS spectra for these
regions. Shown in Fig.~\ref{gsfit} are the results of fits to the GIS
spectra for each region with a single temperature MEKAL model from
XSPEC (with fixed redshift of $z=0.0104$, fixed interstellar
absorption of $n_H=8\times10^{20}$~cm$^{-2}$). The temperature of the
gas, the heavy element abundance, and the normalization were free
parameters.  Note that although the actual shape of the spectrum may
be more complex than the model we are considering (in particular in
the cooling flow region, see e.g. Fukazawa et al.\ 1994), for our
purpose this simple model provides a useful characterization of the
spectral hardness (temperature). Since we are interested primarily in
the temperature variations outside cooling flow regions, a single
temperature approximation is sufficient. This was further confirmed by
values of $\chi^2_r\sim 1$ for the outer regions.

In Fig.~\ref{gsfit}, dotted lines separate the results obtained for
different annuli. NGC4709 falls into region \#24 which has a higher
temperature than adjacent regions. Inspection of other annuli shows
that in general, the temperature is higher in the direction from
NGC4696 towards NGC4709. Comparing Fig.~\ref{gsfit} with the two
dimensional temperature map (Fig.~\ref{gistmap}) one can see that both
methods give consistent results.

\section{EVIDENCE FOR MERGING}

\begin{figure*}
\plottwo{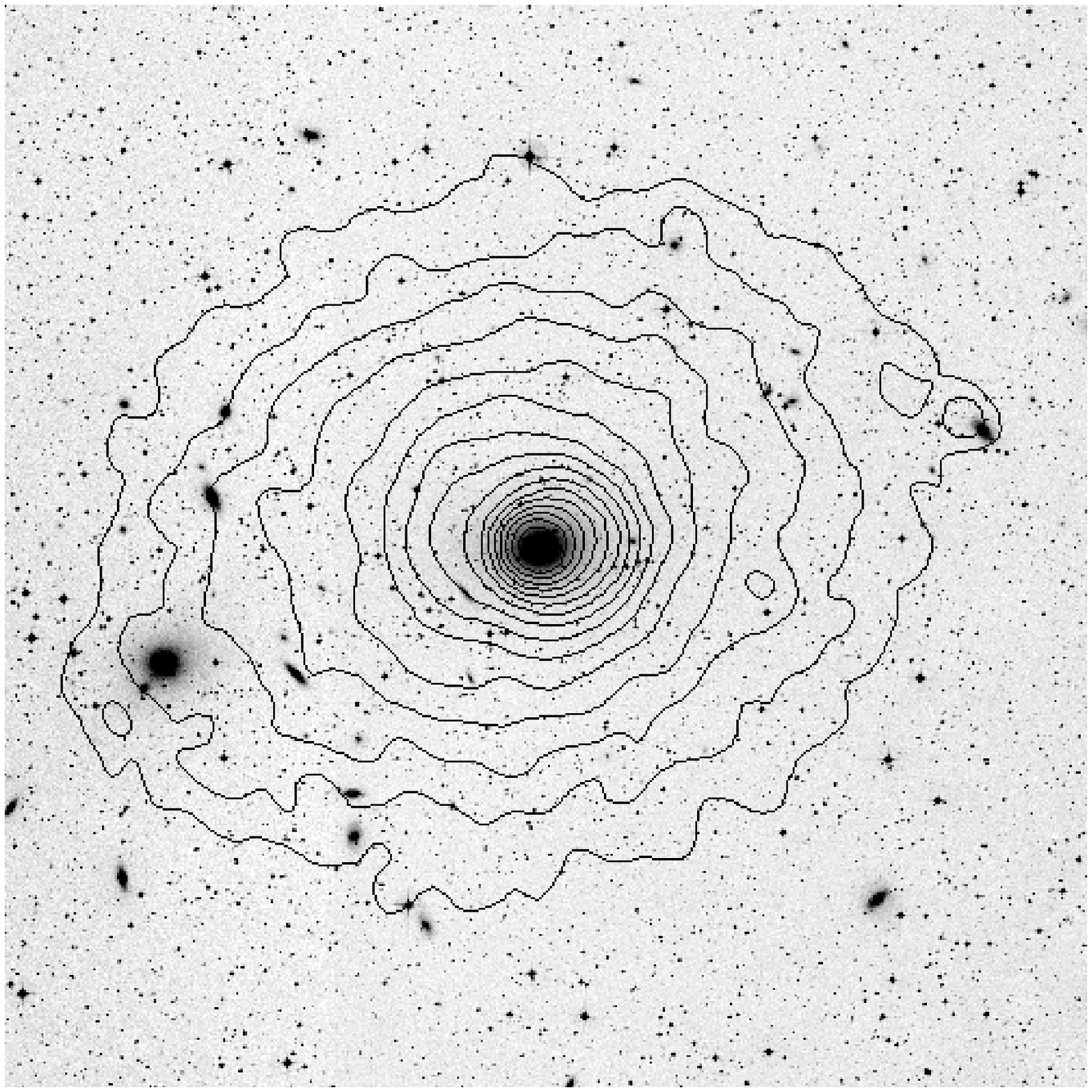}{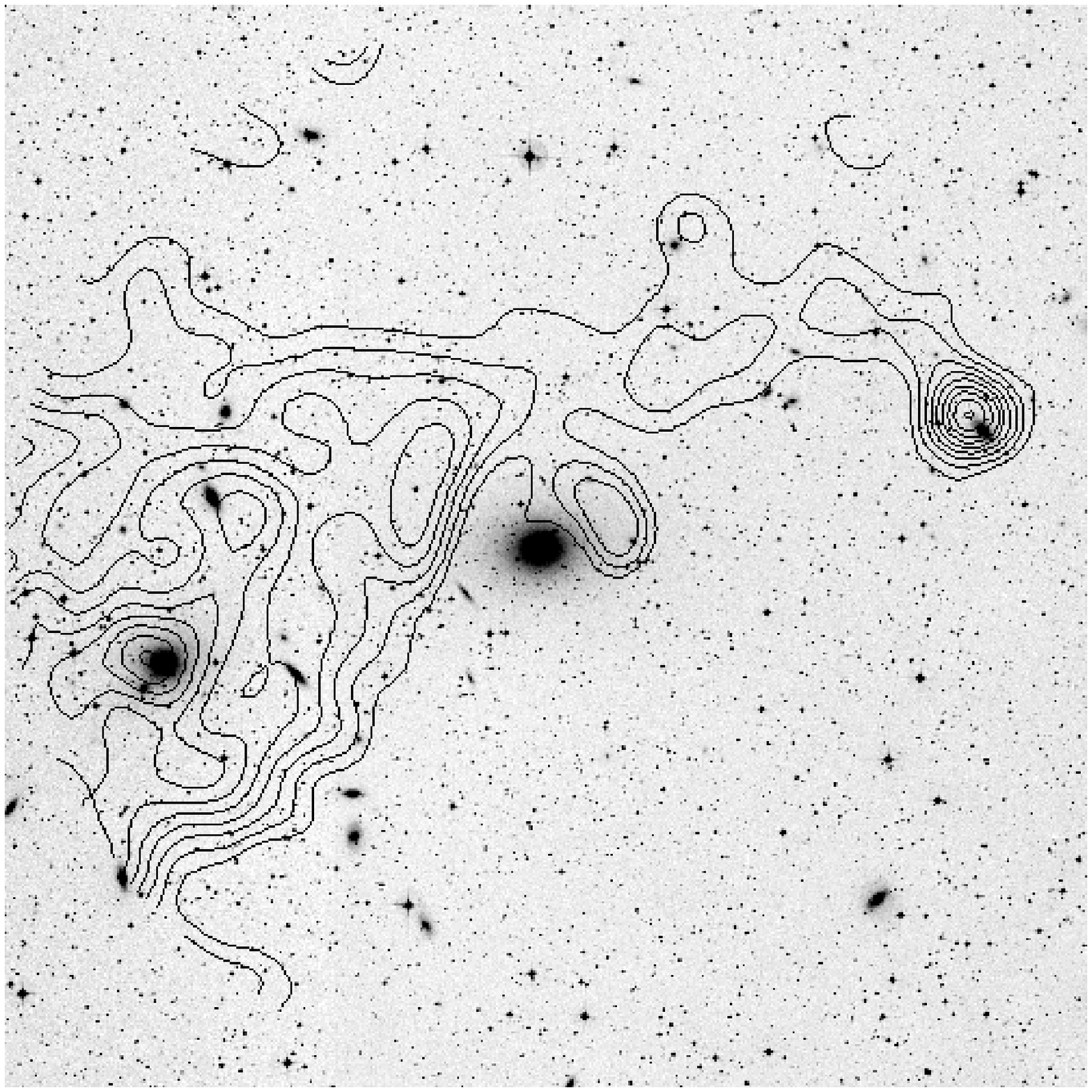}
\caption{{\bf Left:} The central (40 $\times$ 40 arcminutes) region of
the ROSAT PSPC image corrected for vignetting overlayed onto optical
image. The bright elliptical galaxy to the South East is NGC 4709; the
galaxy to the North West is NGC 4696B. The X--ray image was smoothed
with a
$\sigma = 0.6'$ Gaussian.  {\bf Right:} Contour plot of the relative
excess emission with respect to the azimuthally
averaged surface brightness for the PSPC
(i.e. $\frac{Data-Model}{Model}$). The image was smoothed with a
$\sigma=45''$ Gaussian. Contours start at 0.05 with a 0.05
increment. The circle outlines the position of the PSPC support
structure. Contours outside this circle were suppressed in order to
avoid possible spurious features induced by the PSPC support
structure.
\label{wv}
\label{asy}}
\end{figure*}

As is discussed by \cite{af94}, the soft X-ray emission from the
Centaurus cluster is slightly elliptical. They also noted that the
second brightest galaxy NGC4709  is
located along the semimajor axis of the cluster. We co-added three
ROSAT PSPC images, RP800607A01, RP800192N00, and RP800607N00, with
total exposure time of about 18000 seconds. No cleaning was applied to
the data to achieve maximal statistical significance. Exposure maps
were generated for each pointing and also coadded.  The resulting
image (after correction for exposure and smoothing with a $\sigma =
0.6'$ Gaussian) is shown in Fig.\ref{wv}. The elongation of the image and the
variation of the centroid of the X--ray emission on  spatial scales
of a few arcminutes are consistent with those reported by \cite{af94}.
To further emphasize the asymmetric structures in the surface
brightness distribution, the image ($I$) was represented as a sum of
``symmetric'' ($I_s$) and ``asymmetric'' ($I_a$) components, where
$I_s$ was calculated as the average surface brightness in $0.5'$ wide
annuli centered at NGC4696. The asymmetric component is equal to the
difference between the original image and the symmetric component
($I_a=I-I_s$). Shown in Fig.~\ref{asy} is the ratio of the asymmetric
to the symmetric components, $I_a/I_s = (I-I_s) / (I_s)$,
smoothed with a $\sigma=45''$ Gaussian. This image is, in fact, the
relative deviation of the surface brightness from the average value at
a given radius. Although such a decomposition of the surface brightness
distribution into two components is of course somewhat arbitrary,
there are several interesting features apparent in the figure. First,
there is excess emission to the west of NGC4696, which is apparently
responsible for the variation of the centroid of the X--ray isophotes as found
in the analysis of \cite{af94}. Note that in the raw image, the
surface brightness peak coincides nicely with the optical position
of NGC4696 (at least better than $\sim15''$ ).  Second, there are two
very clear excesses of emission. One lies close to the position of the
S0 galaxy NGC4696B (see also Allen \& Fabian 1994) and another is centered on
NGC4709 (which was not detected by \cite{af94} who used cleaned data
from only one pointing having a total exposure time of 3311
seconds). From Fig.~\ref{asy}, it is clear that NGC4709 marks the
center of extended structure, lying to the east
(and south--east) of NGC4696. Given the fact that NGC4709 is the
dominant galaxy in the low velocity dispersion subgroup, Cen 45 (see
Lucey et al. 1986), this suggests that this extended emission is associated
with Cen 45.  Of course, the intensity and extent of the structure
around NGC4709 depends on the assumed centroid position used
for calculation of the ``symmetric'' component. However, the detection of
the excess emission close to NGC4709 is stable against variations of
the centroid position for the symmetric component.

Since the Cen 45 group contains a factor of 2.5 fewer galaxies than
Cen 30 (Lucey et al. 1986) and a significantly lower
velocity dispersion (280\kms~ compared to 586\kms~ for the main
group), a simple projection of the two subgroups would result in an
excess surface brightness around NGC4709 (consistent with the ROSAT
data), but a lower temperature in this region (contrary to the ASCA
data, see Fig.~\ref{gistmap}). We suggest that the most straight
forward explanation would be to assume that the gas in the region
around NGC4709 is heated due to the interaction between the two
subgroups.  Note that the direction of the line joining NGC4709 and
NGC4696 is co-aligned with features at various spatial scales. As
noted by \cite{af94}, the major axes of the elliptical X-ray isophotes
around NGC4696 (at $r \gax 100$ arcsec) are aligned towards NGC4709 as
are the optical isophotes of NGC4696 itself.  At scales up to a few
degrees, this same orientation  marks the chain of separate
groups, identified by \cite{lcd86}. Thus, it may be that the Centaurus
cluster is accreting matter (in rather small increments) along this
preferential direction which perhaps marks the filamentary structure
at large scales (see e.g. West, Jones, \& Forman 1995). At the present
time, the subgroup centered on NGC4709 is merging with the main
cluster.

Another interesting feature which can be seen in Fig.~\ref{wv}
is an X-ray source close to the position of the S0 galaxy
NGC4696B. \cite{af94} noted that this X--ray source probably is
associated with NGC4696B, which is one of the three brightest galaxies
in the field (only NGC4696 and NGC4709 are brighter than NGC4696B in
the optical band). However, the X--ray position is offset
from the optical center of the galaxy by $\sim1'$ towards the
north--east. Moreover, in Fig.~\ref{asy} one can see a long and narrow
filament which originates close to NGC4696B and extends to the
north--east and then to the east. The filament is barely visible in
the raw images and appears clearly only after subtraction of the
component centered on NGC4696. The significance of the excess at
various positions in the feature (defined as the ratio of the signal
at a given position to the statistical error associated with the
smoothed image) corresponds roughly to 2--3 $\sigma$.  Note however
that nearby pixels are not independent on the smoothed image.

Of course it is difficult to prove that the observed excess is a
single feature. One can speculate however on the the origin of this
feature using the same line of arguments as presented by \cite{vfj96},
who discussed a filament of X--ray emission in Coma.  The two simplest
scenarios assume that such a feature may be due to cold gas (ram
pressure stripped) from NGC4696B, like that observed for M86, or due
to a tidally disrupted dark halo from a group of galaxies during its
passage through the central part of the Centaurus cluster. Future
observations (AXAF, XMM and SPECTRUM--X--GAMMA) may distinguish
between these two possibilities: a lower temperature and higher
abundance of the excess emission would suggest stripping, while a lack
of spectral changes would argue in favor of a perturbation in the
gravitational potential.

The shift in the position of the peak of the X--ray emission from
NGC4696B to the northeast of the optical position supports the
suggestion that the whole feature is somehow associated with
NGC4696B. Note also that the major axis of the optical emission of
NGC4696B is also aligned in the direction of the tail. Assuming a
radius for the feature of $\sim$25 kpc, a length $\sim$200 kpc, and an
enhanced electron density of $\sim10^{-3}~{\rm cm}^{-3}$ (required to
produce the observed surface brightness increase), the total gas mass
of the feature is of the order of $6\times10^9~M_\odot$.  The
required stripping rate can be estimated as $16~M_\odot~{{\rm
v}_{500}}~ yr^{-1}$, 
where ${\rm v}_{500}$ is the galaxy velocity in units of 500 km sec$^{-1}$.
This value is too high for steady mass ejection by this single galaxy,
suggesting either that the situation is transient or a group of galaxies is
responsible for this gas. Considering that Centaurus is likely to be
undergoing repeated mergers with smaller subgroups, we can expect
that relatively gas rich galaxies are frequently entering the dense
environment of this cluster.  The evaporation time (assuming that
this is cold gas stripped from NGC4696B) of such a feature due to
thermal conduction would be short ($\sim 10^{7}$ years) and one
would have to assume that thermal conduction is suppressed below
Spitzer's value, which is not unlikely (see e.g. Pistinner, Levinson,
\& Eichler 1996).  As
discussed by \cite{sbs91}, stripped cool gas from cluster galaxies may
fall to the bottom of the potential well and provide strong density
fluctuations in the cooling flow region which are required to trigger
thermal instabilities.  It is interesting that the assumption that the
observed filament arises from cool gas stripped from NGC4696B may
indicate a rather high infall velocity for the cool gas. Indeed, if
the bending of the tail towards NGC4696 is due to infall of the gas in
the cluster potential, then the infall velocity should be comparable
to the velocity of NGC4696B. The infall velocity of an overdense blob
of radius $r$ cannot exceed its terminal velocity ${\rm v}_t\sim {\rm v}_K
\sqrt{(\Delta\rho r)/(\rho R)}$ (e.g. Nulsen 1986), where ${\rm v}_K$ is the
Keplerian velocity and $R$ is the distance from the cluster center. If
most of the mass in the feature is in X--ray emitting gas, then
\begin{eqnarray}
{\rm v}_t\sim {\rm v}_K \left ( \frac{\Delta\rho r}{\rho R} \right )^{1/2}={\rm v}_K  
\left [ \left ( \frac{\Delta\rho^2 r}{\rho^2 R} \right ) \left (
\frac{r}{R} \right ) \right ]^{1/4}
\end{eqnarray}
The $\frac{\Delta\rho^2 r}{\rho^2 R}$ term is the surface brightness
enhancement towards the feature, which is roughly 0.2--0.3, while
$r/R\sim 0.1$. Therefore, ${\rm v}_t\sim 0.4 {\rm v}_K$. Infall of the
blob at this 
velocity may be consistent with the curvature of the feature, if the velocity
of NGC4696B is not much higher than ${\rm v}_K$. Such a high infall velocity
would result in the tail, arising from
stripped gas (which is
visible in the images, i.e. $\frac{\Delta\rho^2 r}{\rho^2 R}$ is not
much less than unity), having one end lying near the cluster center.
However, the tail may be disrupted
as it moves through the cluster gas (e.g. Nulsen 1996), effectively
decreasing the terminal velocity of the infalling gas as it fragments
into smaller blobs. Considering typical
parameters for cooling flow clusters, \cite{sbs91} found an infall
velocity for stable blobs of roughly 60~\kms, assuming a magnetic field
of $\sim 1 \mu G$. Such  low infall velocities would mean
that the shape of the tail is just a reflection of the projection of the galaxy
trajectory (which therefore must have passed through the very center of the
Centaurus cluster).

\section{DISCUSSION}

Despite of the large difference ($\sim 1500$\kms) in the velocities of
the Cen 30 and Cen 45 groups, \cite{lcd86} presented strong arguments
that these two groups are located at the same distance.  Luminosity
functions, color--magnitude relations and galaxian-radius
distributions support the assumption of a common distance and
disagree with the contention that Cen 45 is significantly more
distant than Cen 30, in accordance with their line-of-sight
velocities. Dressler (1995) came to the same conclusion studying
surface brightness fluctuations of several galaxies from both
groups. These results imply that the smaller Cen 45 group is
merging with the dominant Cen 30 group. The analysis of X--ray data
further supports this scenario. Indeed, if one were observing the
projection of two independent clusters, we would expect the
overlapping region to have a lower temperature due to the predicted
lower temperature of the smaller group Cen 45.
Instead, we find that the temperature is
highest in the region close to the dominant galaxy of the Cen 45
group, which we interpret as  heating due to the interaction of the
two subgroups. Note however that the presence of the cooling flow
around NGC4696 restricts
our ability to correctly estimate the extent of the hot region. We might
also expect that when a poor group is entering a dense
cluster environment  for the first time, the relatively cold gas will be
stripped due to the high ram pressure. The marginally significant filament
near NGC4696B, seen in the ROSAT PSPC images, might be explained as
the result of
such an interaction.

Recently \cite{sjf97} found, in the $0.75$\deg circle
around NGC4696, a very large velocity dispersion for Cen 30 ($933\pm
118$\kms) and a  very low dispersion for Cen 45 ($131\pm
43$\kms). The assumption that the main cluster has a velocity dispersion of
933\kms~ would make the so--called $\beta$-problem very
extreme. This parameter (which characterizes the ratio of energy per
unit mass for galaxies to that in the gas) can be estimated from the X--ray
surface brightness distribution, $\beta_f$, or calculated from the observed
velocity dispersion and gas temperature, $\beta_s$. The slope of the
PSPC surface brightness profile implies values of $\beta_f\sim$
0.4--0.45. However, $\beta_s=\frac{\mu m_p \sigma_r^2}{kT}\sim$1.4 for
$T\sim$ 4 keV and $\sigma_r=$933\kms. Even though these values for $\beta$~
 disagree for many clusters, for Centaurus this disagreement is
far larger than usual. An alternative and more likely explanation for
the large velocity dispersion is that
a large scale filament, which defines the preferential
direction of accretion onto the cluster, lies close to the line
of sight. This would explain the high observed velocity
difference of the infalling group and would also account for the
large velocity dispersion observed  by \cite{sjf97}.
The presence of a large scale filament crossing the center of
Cen~30 also would naturally explain the alignments already mentioned
above. Thus, accretion along a filament, lying mostly along the line of
sight, but also partially along the direction defined by the
centers of Cen~30 and Cen~45, would explain the common alignments of 1)
the major axis of NGC4696, 2) the filament emanating from NGC4696B and
3) the nearby chain of groups (Lucey et al. 1986).

The effects of large scale filaments are seen around other clusters.
For example, observations of A1689 suggest the presence of a large
scale filament directly along the line of sight (Daines et al. 1998).
In A1689, the three brightest galaxies in the direction of the cluster lie
within $20''$ of the cluster center, only one has a velocity equal to
the cluster mean while the other two have velocity differences of
+4767 \kms and -2686 \kms, far too large to be explained by the
velocity dispersion of a relaxed $\sim10$ keV temperature cluster.
The nearby cluster A85 also shows the common alignment of several
galaxy concentrations/subclusters with the major axis of the central
cD galaxy (Durret et al. 1997). The X-ray temperature map for A85 (Markevitch
et al. 1998) shows a hot region, suggestive of a shock produced by one
of the infalling subclusters.

Large scale filaments and their effects on clusters also are found in
recent numerical simulations which have adequate resolution.  Colberg
et al. (1998) showed that clusters accrete matter from a few preferred
directions, defined by filamentary structures, and that the accretion
persists over cosmologically long times. Thus, it should not be
surprising to see common alignments between central galaxy major axes,
infalling subclusters, surrounding groups, and, even on the scale of
neighboring clusters (West et al. 1995).  With the
availability of cluster temperature maps, the X-ray observations allow us to
differentiate between superpositions (chance alignments) and true
mergers where shock heating occurs. Hence, as shown in this paper for
Centaurus, we can now study the details of the merger process which
provide strong evidence for accretion along preferred directions as
predicted by numerical simulations of hierarchical structure
formation.

\acknowledgements

This research has made use of data obtained through the High Energy
Astrophysics Science Archive Research Center Online Service, provided
by the NASA/Goddard Space Flight Center. 

%E. Churazov and M. Gilfanov acknowledge support from RFRF grant 93-02-17167.
W. Forman and C. Jones acknowledge  support from NASA contract NAS8-39073.

%\clearpage


\begin{thebibliography}{}

\bibitem[Allen \& Fabian (1994)]{af94} Allen S.W., Fabian A.C., 1994
        \mnras, 269, 409
\bibitem[Churazov et al.\  (1996)]{cgfj96} Churazov E., Gilfanov M.,
        Jones C., Forman W., 1996  
        \apj, 471, 673
\bibitem[]{colberg1998}Colberg, J., White, S.D.M., Jenkins, A., \&
Pearce, F.R., 1998, submitted to MNRAS (astro-ph/9711041).
\bibitem[]{daines98} Daines, S., Jones, C., Forman, W. Tyson, A. 1998,
submitted to ApJ
\bibitem[Dressler (1995)]{dre95} Dressler A., 1995, Mount Stromlo and
                 Siding Spring Observatories,   Heron Island Workshop
                 on Peculiar Velocities in the Universe 
\bibitem[]{durret98} Durret, F., Forman, W., Gerbal, D., Jones, C.,
and Vikhlinin, A., 1998, submitted to A\&A.
\bibitem[Fukazawa et al.\ (1994)]{f94} Fukazawa Y. et al., 1994
        \pasj, 46, L55
\bibitem[Gilfanov et al.\  (1998)]{gcfj98}  Gilfanov M., Churazov E.,
        Jones C., Forman W., 1998   
	(in preparation)
\bibitem[Jones \& Forman (1978)]{jf78} Jones C., Forman W., 1978 
        \apj, 224, 1
\bibitem[Lucey et al. (1986)]{lcd86} Lucey J.R., Currie
        M.J. and Dickens R.J., 1986
        \mnras, 222, 427
\bibitem[Markevitch et al. 1998]{mar98} Markevitch, M., Forman, W., Sarazin,
C., and Vikhlinin, A. 1998, submitted to ApJ
\bibitem[Matilsky, Jones, \& Forman (1985)]{mjf85} Matilsky T., Jones
        C., Forman W., 1985 \apj, 291, 621
\bibitem[Navarro, Frenk \& White (1995)]{nfw94} Navarro J., Frenk C.,
        White S., 1995, \mnras, 275, 720
\bibitem[Nulsen (1986)]{nul86} Nulsen P., 1986, \mnras, 221, 377
\bibitem[Pistinner (1996)]{ps96} Pistinner S., Levinson, A., Eichler,
D. 1996, ApJL, 467, 162
\bibitem[Soker et al. (1991)]{sbs91} Soker N, Bregman J., Sarazin C., 1991
        \apj, 368, 341
\bibitem[Stein, Jerjen, \& Federspiel (1997)]{sjf97} Stein P., Jerjen
        H. and Federspiel M., 1997, astro--ph/9707211
\bibitem[Vikhlinin, Forman, \& Jones (1996)]{vfj96} Vikhlinin A., Forman
         W., Jones C,  1996  
        \apj, 474, L7
\bibitem[West, Jones, \& Forman (1995)]{wjf95} West M., Jones C., Forman
        W., 1995  
        \apj, 451, L5
\bibitem[White et al. (1991)]{whi91} White D.A., Fabian A.C., Jones C., Forman
        W., Stern C,  1991  
        \apj, 375, 35
\bibitem[White, Jones \& Forman (1997)]{white97} White, D.,
         Jones, C., \& Forman, W. 1997
        \mnras, 292, 419
\end{thebibliography}
\end{document}